\documentclass[11pt,a4paper]{article}
\usepackage[english]{babel}

\usepackage[utf8]{inputenc}
\usepackage[T1]{fontenc}

\usepackage[a4paper, top=1in, bottom=1in, left=1.1in, right=1.1in]{geometry}

\usepackage[tagged, highstructure]{accessibility}
\usepackage{graphicx}
\usepackage{authblk}
\usepackage{amsmath,amssymb,amsthm}
\usepackage{url}
\usepackage{pifont}
\usepackage{color}

\newcommand\cparagraph[1]{\vspace{0.6mm}\noindent\textbf{#1.}}
\let\origref\ref
\def\ref#1{\textnormal{\origref{#1}}}

\begin{document}

\title{Democracy in the Era of Artificial Intelligence}

\author[1,2]{Evangelos Pournaras}
\author[1]{Srijoni Majumdar}
\author[3]{Carina Hausladen}
\author[4,5]{Dirk Helbing}

\affil[1]{School of Computer Science, University of Leeds, Leeds, UK. Email: \{e.pournaras,s.majumdar\}@leeds.ac.uk}
\affil[2]{School of Energy Systems, LUT University, Finland}
\affil[3]{Computational Social Science, University of Konstanz, Konstanz, Germany. Email: carina.hausladen@uni-konstanz.de}
\affil[4]{Computational Social Science, ETH Zurich, Zurich, Switzerland. Email: dhelbing@ethz.ch}
\affil[5]{Complexity Science Hub, Vienna, Austria}

\date{}

\maketitle

\begin{abstract}

Interfacing Artificial Intelligence (AI) with democracy is one of the most profound challenges of our times. On the one hand, AI comes with opportunities to overcome long-standing challenges in democracy, such as low participation in deliberative and voting processes with poor representation of people. On the other hand, new risks arise from AI algorithms that are privacy-intrusive, biased, manipulative, spread misinformation and influence election results. Moving beyond the over-simplistic question of whether AI is good or bad for democracy, the Handbook on Democracy in the Era of Artificial Intelligence asks instead: how to upgrade democracies and the principles they are built on, using AI? How to engage with AI and on what terms? Which new values and design principles are required to build democratic resilience? In 34 chapters by 59 authors across the world from different disciplines, we explore how AI can empower collective intelligence for democracy (Part 1) and what is the future of deliberative democracy using large language models and social media (Part 2). We also illustrate the role of AI for building resilient self-governance systems (Part 3) and the challenges of transforming democracy in the age of AI (Part 4). We conclude with broader perspectives (Part 5) that re-imagine the interplay of democracy and AI. 
\end{abstract}

\section*{Introduction}

Democracy has always been a work in progress. From the Athenian Agora to the modern parliament, from the printing press to the Internet, each era has brought new tools that transformed how citizens communicate, deliberate, and organize government. 

Some of these transformations strengthened democratic life; others introduced new vulnerabilities, new concentrations of power, and new opportunities for manipulation. Artificial Intelligence is the latest — and arguably the most influential — of these transformations, and this book is an attempt to understand what it means for democracy today and in the times ahead\cite{metakides2023democracy,ovadya2023reimagining}.

Around the world, democracies are currently under pressure. They are being challenged in multiple ways: by other political systems, by corporations, and by the digital revolution. Among others, democracies are undermined politically by mis- and disinformation, which have even become part of hybrid war, particularly cognitive warfare \cite{deppe2024cognitive}. 

Corporations, on the other hand, have intensely lobbied politics and pushed for major transformations of democratic institutions for the sake of greater efficiency. In favor of profit maximization, this has often undermined fundamental democratic principles such as human rights and human dignity \cite{Helbing2026,jungherr2023artificial}.

Last but not least, Big Tech has often framed democracies as “outdated technology”, which should be replaced by disruptive technological innovations. In fact, the latter have enabled entirely new, data-driven (“cybernetic”) governance models \cite{Helbing_another2026}. Perhaps most notable is the creation of utopian \textit{private} smart cities \cite{gebel2018free}, which have their own organizational form and legislation, instead of a commons-based approach to sharing and managing resources~\cite{Pournaras_another2026,Pournaras2018,Pournaras2020}. 

Therefore, one should discuss the question of why current solutions are still so far behind the promises made, thereby revealing opportunities for improvement \cite{helbing2021next,Helbing2015}, as outlined below.

\section*{Design Challenges for Democratic AI}

Currently, AI systems used to govern societies are facing various challenges:
\begin{itemize}
	\item[(1)]	\textbf{Technical limitations:} Even though we have more data and processing power than ever, data transmission has not kept up with data processing, which has not kept up with data generation, which has also not kept up with the rise in complexity resulting from our increasingly networked systems. Furthermore, while parameter-rich models are easy to calibrate with Big Data (assuming the applied learning algorithms converge sufficiently fast), there are issues with biases, transferability, and interpretability. The often applied black-box algorithms can also not easily explainwhy certain actions have been taken (think of classification errors and hallucinations). 
	\item[(2)]	\textbf{Fundamental limitations:} Real-world applications may face issues such as Gödel’s undecidability problem \cite{godel1992formally}, Turing’s halting (i.e. computability) problem \cite{turing1937computable}, or limits of preditability (“Maxwell’s demon” \cite{rex2017maxwell}, turbulence \cite{leith1972predictability}, deterministic chaos \cite{werndl2009new}, etc.).
	\item[(3)]	\textbf{Optimization:} The solution of some optimization problems is computationally overly demanding due to algorithmic complexity~\cite{Pournaras2018}. Furthermore, there is no established science telling us what goal function to optimize for. It has even been questioned that optimization should be applied to social systems, given that focusing on a one-dimensional goal function is a gross over-simplification, and pluralistic, co-evolutionary approaches can sometimes perform “better than optimal” \cite{pournaras2025upgrading}. 
	\item[(4)]	\textbf{Automation:} While automation works well in areas such as production and logistics, it is less suitable for social systems such as “smart cities”, where human agency matters~\cite{helbing2021next,Pournaras_another2026}. It has also been questioned whether autonomous weapons and data-driven, AI-based systems generally make societies more just, safe, and secure. 
	\item[(5)]	\textbf{Predicability:} Many data-driven applications aim at predictability, and predictability favors control, which may be attempted also by means of behavioral manipulation~\cite{susser2019online,Asikis2021}. While such predictability may produce unilateral profits or power in cases of information asymmetry (for example, enable new kinds of insider trading), it may overall degrade the system performance (e.g., sidelining diplomacy that often succeeds in replacing likely bad solutions by possible better solutions).  
	\item[(6)]	\textbf{Complex dynamics:} Complex dynamical systems (particularly those showing randomness, delays, network effects, and/or extreme events) are often difficult to model, predict and control by machine learning approaches, particularly unsupervised ones~\cite{caldarelli2023role}. Systemic instabilities (“butterfly effects”~\cite{lorenz2000butterfly,palmer2014real}) are illustrative of the challenge.
	\item[(7)]	\textbf{Socio-technical systems:} Cyber-physical systems and Internet-of-Bodies applications have created socio-technical systems that cannot be well understood merely by understanding the technologies involved \cite{sony2020industry,Helbing_another2026}. They show emergent behaviors that call for a systemic perspective considering human and social factors. They also call for social, not just technical innovations. 
	\item[(8)] \textbf{Role of participation:} Today’s networked systems imply that local individual decisions may impact other parts of the system. To turn this into a desirable feature, socio-technical systems should pursue a participatory approach~\cite{mahajan2022participatory}. Deliberation and co-creation are not just integrating previously established knowledge, but produce innovative solutions that go beyond those offered by generative AI alone~\cite{Majumdar2026}. Top-down control approaches undermine such innovation capacity.
	\item[(9)]	\textbf{Positivistic approach:} A data-driven approach overemphasizes measurable quantities and risks. Qualities that humans care about (such as friendship, love, solidarity, trust, freedom, creativity) may be neglected and may get lost in a data-driven society~\cite{pournaras2025upgrading}.
	\item[(10)]	\textbf{Ethics, human rights, and human dignity:} Past technical solutions for social systems have often been quite ignorant about ethics, human rights, and human dignity \cite{helbing2021ethics}, which questions the benevolent, human-centric approach that is often claimed to be the goal of digital innovations. Design for values~\cite{friedman2013value,van2015design,Asikis2021}, particularly for constitutional and cultural values as well as for the values expressed in the UN Charter of Universal Human Rights can help to find improved solutions~\cite{helbing2023democracy}.  
\end{itemize}

\section*{Why a Handbook on Democracy in the Era of AI Is Needed}

Given all these issues, there is an urgent need to consider and explore new ways to upgrade democracies and the principles they are built on, in particular the ability to create and use collective intelligence. Importantly, democracies thrive on pluralism, while the personalization we see in many digital applications is often utilitarian in nature. Now, however, digital technologies, particularly recent AI applications, offer interesting new opportunities for better democracies, which is reflected in the chapters of this handbook.

According to it, the question is not whether AI is good or bad for democracy. This framing is too simplistic. AI is already embedded in the institutions, platforms, and everyday interactions that shape political life. It curates the news we read, moderates the conversations we have online, assists governments in allocating resources, and increasingly informs the decisions of courts, bureaucracies, and electoral systems. Therefore, the question is not whether to engage with AI, but how — and on whose terms. This is addressed by this handbook in five parts, which highlight different areas of research, development, application and policymaking.

\section*{Organization of this Handbook}

\textbf{Part 1: AI-Assisted Collective Intelligence for Democracy} opens this handbook by examining how collective intelligence can enhance the quality of democratic decision-making in real world and how AI can be used in a socially responsible way to foster, coordinate and empower such collective intelligence. The contributors explore fair voting methods, computational approaches to social choice, coordinated resource sharing in Smart cities and digital platforms as living experiments in collective governance. Together, these chapters make the case that AI, when thoughtfully designed and used, can help citizens, cities and institutions aggregate preferences, identify common ground, and strengthen the legitimacy of democratic outcomes.

\textbf{Part 2: The Future of AI-Assisted Deliberative Democracy} turns to the deliberative dimension of democracy — the idea that good decisions emerge not just from aggregating preferences, but also by the very process of a public debate. The chapters here ask how generative AI and Large Language Models (LLMs) can support rather than replace human deliberation, while also confronting the ways in which algorithms, misinformation, and platform design can derail it. The tension between scaling deliberation and preserving its authenticity runs throughout.

\textbf{Part 3: AI for Resilient Self-Governance Systems} shifts the attention to decentralized governance — communities, commons, and autonomous organizations that are experimenting with AI-enabled coordination outside traditional institutional frameworks. These chapters explore how self-organization can outperform top-down crisis response and what the rise of decentralized autonomous organizations means for democratic accountability and human agency.

\textbf{Part 4: Challenges for Transforming Democracy in the Age of AI} confronts the harder questions head-on. Parliaments, governments, and regulators are struggling to keep pace with technologies that advance faster than the law. Surveillance, profiling, and targeting threaten the privacy that democratic participation depends upon. The European Union's regulatory attempt offers one approach; the chapters here assess its ambitions and its limits, alongside a broader reckoning with what meaningful democratic oversight of AI actually means.

\textbf{Part 5: Reimagining the Interplay of Democracy and AI} steps back to reflect on the deeper philosophical and cultural issues. What does AI mean for the foundations of democracies and the concepts they are built on? How do media narratives shape the public understanding of these technologies? And what are the long-term risks — to education, the functioning of our minds, and what it means to be human — in a world increasingly organized by artificial intelligence?

All five parts share the vision that the relationship between AI and democracy is not predetermined. Technology does not have the will of people built in by default, but it can be built in by design. If AI is reshaping the very conditions of democratic life, then it is an important task of citizens, scholars, policymakers, and algorithm designers to ensure that this reshaping is itself subject to democratic principles. Accordingly, the chapters that follow offer rigorous, pluralistic, and open-minded perspectives on core challenges of our time.

The rest of this introductory chapter outlines the contributions in each of the five parts of the handbook.

\section*{Part 1: AI-assisted Collective Intelligence for Democracy}\label{sec:collective-intelligence}

This opening part of the handbook hosts the following six chapters: \\

	\cparagraph{2. Fair Voting Methods as a Catalyst for Democratic Resilience: \\A Trilogy on Legitimacy, Impact and AI Safeguarding} \\ 
	Drawing on the growing momentum of adopting fairer voting methods in the city of Aarau in Switzerland as well as other cities in Poland and across the world, Pournaras~\cite{Pournaras2026} highlights strong empirical evidence of the transformative character of fair voting methods: more winners with higher voters' representation, improved spatial fairness, stronger legitimacy based on democratic values such as compromise and altruism. Citizens' decisions come with stronger impact on welfare, education and culture, while biases of AI voting assistance are mitigated. Pournaras~\cite{Pournaras2026} also discusses how fairer voting methods can build higher resilience for democracies in deep crisis such as that of Greece. \\

	\cparagraph{3. AI and Computational Social Choice}\\ 
	Talmon et al. \cite{Talmon2026} bring a timely perspective on the developing relationship between democracy, computational social choice and AI, showing how AI can reshape the foundations of collective decision making and democratic governance. The chapter demonstrates how the use of AI is relevant across the full decision-making pipeline:  assisting with incomplete preference elicitation, supporting proxy voting and liquid democracy through intelligent delegation, automating the design of voting rules that reflect historical patterns, community satisfaction, and human values. Three key decision-making aspects are analyzed: (i) AI as a tool for richer and more adaptive preference elicitation, (ii) AI as an enabler of collaborative and similarity-based vote delegation, and (iii) AI as a methodological framework for designing voting rules. \\
	
	\cparagraph{4. Digital Democracy as a Living Experiment} \\
	Yang \cite{Yang2026} presents democracy not as a finished institution but as a living experiment, constantly tested, revised, and redesigned as societies change. The chapter is divided into five sections, offering a comprehensive review and analysis that begins with the historical and mathematical critiques of democratic theory, including Arrow’s impossibility theorem, and then proceeds through pluralism, participation, computational social choice, and digital democracy. The chapter’s central provocation is powerful: Democracies survive when they are treated as evolving systems, capable of being refactored, repaired, and boldly reimagined when existing designs no longer serve the public good. \\
	
	\cparagraph{5. Beyond Human Mediation: \\AI, Autonomous Data Perception, and the Future of Communication and Democracy} \\
	Ahn et al. \cite{Ahn2026} invite readers to rethink democracy in an age where algorithms do not just support decisions, but increasingly shape what people see, know, and believe. The central message is clear and urgent: democracy must remain a human practice of judgment, disagreement, responsibility, and shared decision-making. Rather than rejecting AI, the chapter analyzes how societies can redesign democratic institutions so that technology strengthens people’s power to think, choose, question, and act for themselves. Three interrelated pathways are investigated and explored for the further evolution of democracy: (i) democratizing algorithmic power by subjecting algorithms to democratic oversight, (ii) designing hybrid knowledge ecologies that supplement human cognition within participatory frameworks, and (iii) cultivating digital civic literacy by fostering algorithmic awareness and critical data thinking. \\
	
	\cparagraph{6. Repairing, Reviving, and Upgrading Democracies in the Age of AI} \\
	Helbing et al. \cite{Helbing2026} present democracy not as a system to be merely preserved, but as one that must be repaired, revived, and upgraded as societies face new challenges in the digital era. This chapter offers a broader and more complete democratic framework: one that is generic enough to be applied across multiple democratic settings and is not limited to simply using AI for democracy. Drawing on Dahl’s view of democracy as an evolving arrangement, it offers a threefold conceptual lens for democratic transformation: (i) repair, to restore core safeguards such as transparency, the rule of law, and checks and balances; (ii) revive, to strengthen weakened practices such as public deliberation, civic trust, and proportional representation; and (iii) upgrade, to integrate digitally supported formats such as participatory budgeting, new voting methods, co-creation, algorithmically supported deliberation, and large language models. \\
		
	\cparagraph{7. \textsc{Another Intelligence}: \\Distributed Socio-technical Systems to Trust for a Sustainable and Democratic Society} \\
	Drawing on his recent inaugural lecture, Pournaras \cite{Pournaras_another2026} introduces his vision for another intelligence: a trustworthy distributed intelligence for a sustainable and democratic society. Through a journey of four career milestones with real-world impact, Pournaras \cite{Pournaras_another2026} unravels how to build socially responsible AI assistance for managing complex distributed socio-technical systems. These include tackling some timely but also long-standing problems, such as helping people buy more sustainable products, coordinating resource sharing in Smart Cities, and upgrading democracies via fair and secure voting systems. Hard questions about the future of trust within a co-evolving human-artificial collective intelligence are discussed. Pournaras \cite{Pournaras_another2026} also shares an interesting story about the role of the young generation in shaping the future of trustworthy distributed intelligence.

	%%%%%%%%%%%%%%%%%%%
	
\section*{Part 2: The Future of AI-assisted Deliberative Democracy}\label{sec:deliberative-democracy}

The following seven chapters constitute the second part of this book: \\
	
	\cparagraph{8. Informed Decision Making in Deliberative Democracy: \\A Generative Artificial Intelligence Framework} \\
	Karacapilidis et al. \cite{Karacapilidis2026} reimagine human deliberation not as a noisy competition of opinions, but as a structured process of collective reasoning, empowering citizens not just to speak, but to question, understand, and shape better democratic decisions. They develop an intelligent assistant for democratic deliberation using a neuro symbolic AI architecture that combines large language models, explainable AI, and knowledge graphs. Deliberation support is investigated through: (i) fact checking, (ii) feedback summarization, (iii) contradiction detection, (iv) argument building, and (v) explanation generation.  \\
	
	\cparagraph{9. Benchmarking AI's Deliberative Reasoning: \\Evaluating LLMs Against Human Collective Wisdom} \\
	Veri et al. \cite{Veri2026} offer a different and unique perspective on AI in deliberative democracy by asking a simple but important question: Can LLM agents actually reason slowly and responsibly enough to support democratic deliberation? They consider the possibility that these models can support a stable set of common reasons and answer with authentic reasonableness, rather than as a tool that simply produces answers. Two key capacities are explored: (i) internal consistency, the ability of an LLM to hold a coherent view of common arguments made during a discussion; and (ii) reasonableness, the ability to weigh trade-offs, anticipate counterarguments, and justify conclusions in a way that aligns with human deliberators. \\

	\cparagraph{10. Human/AI Collective Intelligence for Deliberative Democracy: \\A Human-Centered Design Approach}\\
	De Liddo et al. \cite{DeLiddo2026} situate deliberative democracy in the practical world of human and artificial collective intelligence and ask how democratic systems can really think, remember, and reason together when many people and machines interact. They argue that artificial intelligence should not replace human judgment but rather assist citizens, institutions, and civic stakeholders in organizing dialogue, discourse, and deliberation in a more transparent, inclusive, and scalable way.  Bringing together democracy, collective intelligence, and deliberative democracy, they show how collective intelligence for democracy can support (i) collective memory and attention, by helping communities track arguments, concerns, evidence, and unresolved tensions over time; and (ii) collective reasoning and accountability, by making disagreements, trade-offs, and decision pathways more visible. \\
	
	\cparagraph{11. The Almost Intelligent Revolution: \\Options for Scaling Up Deliberation and Empowering People with AI} \\
	Sharoff \cite{Sharoff2026} reflects that democratic deliberation often works in favor of people who already know the right vocabulary, style, and forms of argument, while others may find it difficult to translate their lived experience into language that is recognized in democratic debate. Sharoff \cite{Sharoff2026} investigates how artificial intelligence can address this by: (i) making complex public and institutional language more accessible to a wider audience, enabling them to understand and respond to it; and (ii) assisting citizens in expressing personal, everyday experiences as clear arguments that can be introduced into public debate. The author adds a necessary note of caution regarding the possibility of bias or false information produced by artificial intelligence, and encourages users to be aware of these risks when using such systems. \\
	
	\cparagraph{12. Mapping the Entangled Landscape of AI and Misinformation} \\
	Stockinger \cite{Stockinger2026} explains why misinformation has become one of the most urgent challenges for deliberative democracy in the age of artificial intelligence. The chapter shows that democratic discussion depends on a shared information environment, but this environment can be weakened by false content, biased systems, deepfakes, social bots, personalized feeds, and automated moderation. At the same time, the chapter avoids a one-sided view of technology by showing how artificial intelligence can also help protect democratic debate through: (i) fact checking, misinformation detection, and content provenance tools that help citizens know what is reliable; and (ii) better governance, digital literacy, and international cooperation that make information systems more accountable. \\
	
	\cparagraph{13. Democracy in the AI Era: YouTube Algorithms and Social Polarization}\\
	Cha \cite{Cha2026} examines how YouTube recommendation systems can shape what citizens see, believe, and discuss, and why this matters for democracy. Cha \cite{Cha2026} explores the potential of community engagement, while also showing how it can deepen political divides. The democratic impacts of YouTube are studied from two different perspectives: (i) how recommendation systems are technically designed and what they are optimized to promote; and (ii) how users, institutions, and political cultures respond to those systems. Its central message is that digital curation can support public discussion, but only if platforms, institutions, and citizens work together to reduce polarization and protect a shared democratic reality. \\
	
	\cparagraph{14. Deliberation Derailed: Brexit, X and the Crisis of Democratic Discourse} \\
	Wellings \cite{Wellings2026} examines yet another social media platform, X, formerly Twitter, to explore how it shaped democratic discussion during the Brexit period from 2016 to 2020. The chapter examines how Twitter fared as a platform for reasoned, inclusive, and evidence based deliberation, and explores (i) how social media can intensify division by rewarding conflict, slogans, and group identity; and (ii) how information bubbles can weaken the shared reasoning needed for deliberative democracy. \\

\section*{Part 3: AI for Resilient Self-Governance Systems}\label{sec:self-governance}

In this third part of this handbook, the following six chapters are illustrated: \\

\cparagraph{15. Power, Empowerment and Power-Sensitive Design in Self-governing Systems} \\
Pitt et. al. \cite{Pitt2026} confront the AI-driven concentration of power in digital democracy, arguing that the unchecked rise of Big Tech platforms is steadily disempowering and entrapping citizens. They propose a new way to give people more power in digital democracy, built around three ideas: better feedback between citizens and institutions, more ethical digital platforms, and AI systems that learn and reason in socially responsible ways. Using power-sensitive design and shared digital governance, the authors show how communities can work with AI to shape and improve their own social systems, rather than simply being controlled by it.  \\

\cparagraph{16. Decentralized Commons-based Governance in the Era of Cyber-Physical Systems} \\
Nardelli et al. \cite{Nardelli2026} argue that artificial intelligence and cyber physical systems can support democracy when they are grounded in shared resources, inclusion, and public participation by design. The authors explore the possibility of these technologies helping communities define goals, evaluate outcomes, and guide decisions together. They present artificial intelligence as a tool for commons-based governance, where technology does not control communities, but helps communities control the systems that shape everyday life. \\

\cparagraph{17. Artificial Intelligence for Democratic Resilience: \\How Self-organization can Outperform Traditional Crisis Response} \\
Banerjee et al. \cite{Banerjee2026} explore a different aspect related to community crises and AI-enabled democratic resilience, showing how artificial intelligence can strengthen democratic resilience during these crises, when top-down institutional responses often fail. They investigate how artificial intelligence can amplify grassroots knowledge, coordinate volunteer networks, and aggregate citizen preferences more fairly than traditional voting, while also showing how it can entrench bias, surveillance, and exclusion if poorly designed. A value-based design approach is studied, grounded in inclusion, transparency, and participatory fairness, arguing that democratic resilience emerges not from institutions alone, but from the dynamic interplay of citizens, technologies, and shared values. \\

\cparagraph{18. Slaying the Dragon: \\The Quest for Democracy in Decentralized Autonomous Organizations (DAOs)} \\
Balietti et al. \cite{Balietti2026} reflect on how blockchain-based organizations aim to make decisions more open, shared, and transparent, while offering ways to build more resilient and accountable AI systems and address the problem of democratic power being centralized. Two potential aspects are investigated: (i) token-based voting can give too much power to a small number of wealthy users, which weakens the idea of shared decision making; and (ii) important technical choices are often still made by small insider groups, so these systems are not as autonomous as they claim to be. The main message is clear: without fairer governance, Decentralized Autonomous Organizations may become blockchain versions of traditional corporations, rather than real experiments in digital democracy. \\

\cparagraph{19. DAO-enabled Decentralized Physical AI: A New Paradigm for Human-Machine Collaboration} \\ Ballandies et. al. \cite{Ballandies2026} propose a new democratic architecture for the governance of physical and digital systems based on coordination principles of humans and autonomous systems. They introduce the concept of DePAI, DAO-enabled decentralized Physical AI that relies on Decentralized Autonomous Organizations. They envision a bottom-up and more democratic way to grow, operate and manage physical infrastructures by communities with the support of AI and blockchain technology to align community needs with security and incentives. Ballandies et. al. see on DePAI an ``optimistic governance" paradigm that balances efficiency with autonomy based on AI agent micro-decisions with human oversight. \\

\cparagraph{20. How Emerging Nano-, Neuro- and Quantum Technologies \\Enable Emerging Cybernetic Societies} \\
Helbing \cite{Helbing_another2026} offers a timely perspective on the shift from data driven societies to emerging cybernetic societies, where AI and digital twins increasingly shape how societies are understood, managed, and governed. The pursuit of safety and sustainability through cybernetic governance may come at a profound human cost, weakening privacy, autonomy, dignity, and the freedom to decide for oneself. Helbing’s central message is that societies must openly discuss these technological futures before  technology-driven lock-in effects and algorithmic control make it very difficult to choose alternative paths. Three key aspects are investigated: (i) the rise of data-driven applications and digital twins, (ii) the shift from human-led governance to cybernetic control systems, and (iii) the need for transparency, democratic debate, and human choice in shaping the future of AI-governed societies. \\

\section*{Part 4: Challenges for Transforming Democracy in the Age of AI}\label{sec:challenges}

The following seven chapters are introduced in the fourth part of this handbook: \\
	
	\cparagraph{21. Reimagining Parliament in the Era of Artificial Intelligence} \\ 
	Fitsilis et al. \cite{Fitsilis2026} discuss the democratic challenge due to the growing disconnect citizens feel from parliamentary functioning and propose an AI integration framework, alongside a five-stage maturity model, for how AI can transform parliamentary activities to become more engaging and informative for citizens. The study identifies seven key domains of parliamentary-side activities, such as: (i) cultural preservation, (ii) innovation, and (iii) diplomatic relations, while it demonstrates how AI can transform each domain. The findings suggest that AI-enhanced parliamentary activities can bridge the gap between institutions and citizens, making parliamentary engagement more interactive rather than a passive transfer of information. \\
	
	\noindent\textbf{22. Governments as Adopters and Regulators of AI: \\A Challenge for Democracy?} \\
	Trein et al. \cite{Trein2026} study how governments can address democratic challenges while managing the dual task of regulating and adopting AI. The adoption of AI in public administration can improve efficiency and make services more responsive, but it also raises concerns about transparency, accountability, and fairness. The chapter shows that governments must not only use AI to modernize public administration, but also govern it carefully so that public trust and democratic safeguards are protected.  The authors examine this through: (i) how AI is framed as a policy problem, (ii) the regulatory approaches adopted in different political systems, and (iii) the politicization of AI governance. \\
	
	\cparagraph{23. Tracking, Profiling, and Targeting: Exploring Misuse Cases of AI} \\
	Zufferey et al. \cite{Zufferey2026} explore a challenge that raises an unsettling question for democracy: Does AI know a voter better than the voters know themselves? The chapter examines how personal data is harvested on a large scale and how AI has amplified this phenomenon. It discusses sophisticated data processing techniques used to build detailed digital profiles of individuals, as well as methods used to influence, predict, and target individual behavior. The authors also offer a hopeful angle by presenting countermeasures aimed at protecting personal privacy in the era of AI. \\
	
	\cparagraph{24. Privacy as a Safeguard for Democracy in the Era of AI} \\
	Longin et al. \cite{Longin2026} examine how personal data are collected from individuals, sometimes with consent, sometimes without consent, and sometimes through pressure or incentives, such as when people are asked to accept cookies to access a service. The chapter explains how AI can threaten both individual privacy and collective privacy by enabling large-scale tracking, profiling, and surveillance. It also shows why this matters for democracy: when citizens are constantly monitored or targeted, their freedom, trust, and ability to participate openly can be weakened. The authors also explore ways to protect privacy through democratic AI governance, stronger safeguards, and greater citizen control over data. \\
	
	\cparagraph{25. Regulating for Democracy: The EU's Experiment with Artificial Intelligence} \\
	Markis et al. \cite{Makris2026} examine whether the European Union’s Artificial Intelligence Act protects against challenges to democracy or unintentionally creates new power imbalances. They argue that the Act is an important step toward better AI governance, but it may not go far enough in addressing deeper democratic concerns. Three key issues are discussed: (i) the Act creates useful rules for compliance and accountability; (ii) it focuses too much on technical and expert-led forms of governance; and (iii) it gives limited attention to the power of private companies and the role of citizen participation. The main message is that AI regulation must protect democracy not only through rules and checks, but also by addressing power, participation, and future risks. \\
	
	\cparagraph{26. AI Strengthening Democratic Engagement: \\Exploring Existing Digital Tools for Public Participation and Future Directions} \\
	Fontes et al. \cite{Fontes2026} continue the theme of tackling low democratic engagement by examining how public participation, aided by artificial intelligence, can become more effective, accessible, inclusive, and meaningful. They review existing digital tools for public participation and identify the main gaps that still prevent wider citizen engagement. They show how artificial intelligence can help by: (i) making participation easier for more people through better access, support, and communication; (ii) helping institutions process large amounts of citizen input more clearly and fairly; and (iii) creating more meaningful ways for people to contribute to public decisions. Fontes et al. also point to future directions for using artificial intelligence to strengthen public participation without losing the human voice of democracy. \\
	
	\cparagraph{27. Decoding Success: Insights from the Consul Democracy Community} \\
	Strohmenger \cite{Strohmenger2026} draws on ten years of experience with the international Consul Democracy community to show how digital citizen participation can help address the challenges democracies face today, from declining trust to weak public engagement. Strohmenger \cite{Strohmenger2026} then shows how artificial intelligence can help scale participation through tools for grouping ideas, translation, moderation, and proposal support, without replacing the foundations of citizen voice and trust. With a careful investigation of risks such as black box bias, power shifts, and AI systems talking only to themselves, Strohmenger \cite{Strohmenger2026} makes a compelling case for AI as an enabler of democratic engagement. \\

%%%%%%%%%%%%%%%%%

\section*{Part 5: Reimagining the Interplay of Democracy and AI}\label{sec:challenges}

The last part of this handbook comes with the following seven chapters: \\
	
	\cparagraph{28. Democracy and Technocracy in the Age of AI} \\
	Bertsou et al. \cite{Bertsou2026} explain technocracy as decision making led by experts, and use it to show how AI and algorithms may shift democratic decisions away from citizens and elected representatives. They compare human expert-led decision making with AI-based decision making, showing that both promise efficient and evidence based outcomes, but AI can make public decisions harder to understand, question, and hold accountable. The authors investigate how citizens judge the legitimacy of AI in public services, especially when trust, fairness, and human oversight are at stake. This chapter encourages readers to consider how AI can be used in democratic systems without weakening citizen voice, responsibility, and public control. \\
		
	\cparagraph{29. AI and Democratic Backsliding: Navigating the Risks of Algorithmic Politics} \\
	Yardımcı \cite{Yardimci2026} explores how AI can dilute democracy when it is used without strict rules, transparency, and public oversight. The chapter shows how AI can support political manipulation through targeted campaign messages, deepfakes, and disinformation, while also making public decision making less clear in sensitive areas such as welfare, policing, and immigration. It argues that these risks are serious, but not unavoidable, as regulation, fact checking, human oversight, and AI literacy can help protect democratic institutions. Yardımcı \cite{Yardimci2026} presents a different view, encouraging readers to consider not only what AI may do to democracy, but also what democracies choose to do with AI. \\
	
	\cparagraph{30. The Impact of AI on the Philosophical Foundations of Democracy} \\
	Hofweber \cite{Hofweber2026} argues that AI changes the way we think about the philosophical foundations of democracy. He shows that AI unsettles democracy’s old foundations: voting, rational debate, and legitimacy become harder to defend when AI can predict what people want, shape what they hear, and influence how they decide together. Hofweber \cite{Hofweber2026} argues that we need to rethink and strengthen the reasons for democracy in an AI age, so that democracy can still be trusted and defended when AI shapes what people know, want, and decide. \\
	
	\cparagraph{31. Technology, Democracy, and Media Art} \\ Naveau et. al. \cite{Naveau2026} bring a perspective from media arts, arguing that arts are inherently political. The key role of arts for the critique of big tech and surveillance capitalism is discussed along with how art can stimulate polyphonic democratic processes contributing to an open society, or how arts can make creative advancements of participatory approaches. Three perspectives are unfolded: (i) arts as a vehicle for democratic values, (ii) democracy as a framework for art and (iii) art as a means for social and political change. \\
	
	\cparagraph{32. Empowering Collectives and Enhancing Decision-Making: \\How Digital Technology Can Address the Crisis of Democracy} \\
	Adams \cite{Adams2026} explains how the weakening of democratic social structures has pushed people to form their identities and interests online, where social media often encourages anger, polarization, and fragmentation. He presents digitally enhanced participatory democracy as a way to restore democratic legitimacy in the age of platform capitalism, by making institutions more responsive to today’s diverse and shifting public interests. Adams provides an approach for evaluating digital democratic tools based on: (i) the empowerment of individuals and communities, (ii) the strengthening of community engagement and democratic relationships, and (iii) the ability to support faster and better decision making by helping people understand complex issues and learn together. \\
	
	\cparagraph{33. Framing Artificial Intelligence: Media Narratives and Public Discourse}\\
	Hänggli et al. \cite{Haenggli2026} examine how AI is framed in the news media and how it shapes public understanding of its role in society. Their analysis shows a sharp rise in AI coverage after 2022, with journalists doing most of the framing, followed by industry actors, while political voices remain at the margins of the debate. They find that AI is mainly presented through progress and economic consequences, often with positive and market-oriented future prospects, while ethics and morality receive much less attention. Hänggli et al. \cite{Haenggli2026} call for a more inclusive and balanced public discourse, where civil society, academia, law and ethics-oriented voices help ensure that AI development serves democratic values, social good and a fair digital future. \\
	
	\cparagraph{34. The AI-Centered Enveloping of the World: \\Overlooked Long-Term Risks to Education, Philosophy and Democracy}\\ Perperidis \cite{Perperidis2026} examines the hidden long-term risks of AI and how they may gradually weaken democracy. The chapter explains that AI may quietly reshape human activities so they fit what machines can handle: education may become just information delivery, philosophy may become only practical problem solving, and democracy may be reduced to simply choosing representatives. Perperidis \cite{Perperidis2026} explores solutions through stronger ethical, legal, and cultural safeguards that can prevent AI from reducing democratic life to algorithmic optimization, while preserving democracy as an open, bottom up process of shared meaning making and collective self-determination. \\

	 \section*{A Democratic Perspective on Democracy in the Era of AI}\label{sec:diversity}
	 
	 The chapters outlined above highlight the comprehensive portfolio of diverse, yet complementary contributions to this handbook. We, 4 editors, brought together 34 chapters by 59 authors from 33 institutions, 12 countries, 4 continents and with a strong representation of different disciplines including among others: history and cultural studies, arts and media, computer science, law, political science, economics, philosophy, engineering and health. Beyond academia, the contributors of this handbook also represent open-source software communities for democratic participation (i.e. CONSUL) as well as democratic institutions and organizations (i.e. parliaments).
	 
	 Understanding the complex interplay of democracy and AI requires diverse knowledge and multi-faceted skills that no single discipline can offer right now. All chapters remain accessible to the broader readers without heavy jargon and as such we encourage readers to navigate through current state-of-the-art at the interface of disciplines for inspiration and exploring out-of-the-box solutions.

	\section*{Acknowledgments}
	
	This handbook has been supported by a UKRI Future Leaders Fellowship (MR\-/W009560\-/1): `\emph{Digitally Assisted Collective Governance of Smart City Commons--ARTIO}', the School of Computer Science at University of Leeds, and the Professorship of Computational Social Science at ETH Zurich. 
	
\bibliography{sample}%bibliography}
\bibliographystyle{naturemag} %apalike} %naturemag}
	
\end{document}